\documentclass[11pt]{article}%
\usepackage{amssymb}
\usepackage{amsfonts}
\usepackage{amsmath}
\usepackage{graphicx}%
\providecommand{\U}[1]{\protect\rule{.1in}{.1in}}

\begin{document}

\title{Improving the\ proof of the Born rule using\ a physical requirement on the
dynamics of quantum particles}
\author{Yakir Aharonov$^{a,b}$ and Tomer Shushi$^{c}$\\$^{a}$Schmid College of Science, Chapman University, Orange\\CA 92866, USA\\$^{b}$Raymond and Beverly Sackler School of Physics\\and Astronomy Tel-Aviv University, Tel-Aviv\\69978, Israel\\$^{c}$Center for Quantum Science and Technology, \&\\Department of Business Administration, Guilford\\Glazer Faculty of Business and Management,\\Ben-Gurion University of the Negev, Beer-Sheva, Israel}
\maketitle

\begin{abstract}
We propose a complete proof of the Born rule using an additional postulate
stating that for a short\ enough time $\Delta t$ between two measurements, a
property of a particle will keep its values fixed. This dynamical postulate
allows us to produce the Born rule in its explicit form by improving the
result given in [1].\ While the proposed postulate is still not part of the
quantum mechanics postulates, every experiment obeys it, and it can not be
deduced using the standard postulates of quantum mechanics.

\end{abstract}

The Born rule is a Postulate of quantum mechanics that provides the structure
of the probability theory of quantum systems, and thus, it plays a key role in
both theoretical and experimental studies of quantum systems. Over the years,
several attempts have been made to prove, or at least trace, the mechanism
behind the Born rule [1-7]. While we do not review recent results in the
direction of proving the Born rule, we point out that a recent paper that
reviews such a topic is given by Vaidman in [2]. In [1], the Born rule was
proved\ by following an assumption about the stability of the system against
small perturbations where it holds only for \textit{likely} outcomes and is
violated for \textit{unlikely} outcomes, such as the case of a large coherent
interference effect between the amplitudes of $N>>1$ decoupled particles in
the system. The proposed proof improves the one given in [1] by imposing a
dynamical physical requirement that holds for every quantum system and for
\textit{likely }as well as \textit{unlikely} outcomes. We start with a proof
of the Born rule using only mathematical arguments, and then we show how an
assumption that is made during the proof can be reformulated into a physical
requirement about the dynamics of the quantum system.

Let\ $\left\vert \psi\right\rangle =\sum_{j}b_{j}\left\vert a_{j}\right\rangle
$ be a prepared state of a particle in the form of a superposition of
non-degenerate eigenstates of some Hermitian operator $A.$ Taking a sample of
$N$ identically prepared nonentangled particles in the state $\left\vert
\psi\right\rangle ,$ the state of this sample is given by the product state%
\begin{equation}
\left\vert \Psi_{\text{total}}\right\rangle =%
{\displaystyle\prod\limits_{i=1}^{N}}
\left\vert \psi\right\rangle _{i}. \label{Psi1}%
\end{equation}
We follow standard quantum theory where if $A$ is measured on state
$\left\vert \psi\right\rangle $, the measurement outcomes are the eigenvalues
$\alpha_{j},$ however, their probabilities $p_{j}$\ are left unspecified, and
the standard Born rule $p_{j}=\left\vert b_{j}\right\vert ^{2}$ will be shown
to uniquely follow from the proposed Postulate.

In [8], the authors proposed the following\ universal formula for $A$ acting
on $\left\vert \psi\right\rangle ,$
\begin{equation}
A\left\vert \psi\right\rangle =\overline{A}\left\vert \psi\right\rangle
+\Delta A\left\vert \psi_{\bot}\right\rangle , \label{formula}%
\end{equation}
where $\left\vert \psi_{\bot}\right\rangle $ is a state vector in the
perpendicular Hilbert space that satisfies $\left\langle \psi|\psi_{\bot
}\right\rangle =0$, with the mathematical quantities%
\begin{equation}
\overline{A}:=\left\langle \psi\right\vert A\left\vert \psi\right\rangle
\equiv\sum_{j}\left(  b_{j}^{\ast}b_{j}\right)  \alpha_{j}, \label{A1}%
\end{equation}
and%
\begin{equation}
\Delta A:=\sqrt{\left\langle \psi\right\vert \left(  A-\overline{A}\right)
^{2}\left\vert \psi\right\rangle }. \label{A2}%
\end{equation}
It is important to mention that both (\ref{A1}) and (\ref{A2}) do not have any
statistical interpretation, and they are merely algebraic operations between
$A$ and $\left\vert \psi\right\rangle .$ The formula (\ref{formula}) can be
proved in various ways, algebraically or geometrically. A simple proof of
(\ref{formula}) runs as follows: For an operator $A$ acting on a vector
state~$\left\vert \psi\right\rangle $ one can write $A\left\vert
\psi\right\rangle =c_{1}\left\vert \psi\right\rangle +c_{2}\left\vert
\psi_{\bot}\right\rangle $ for some $c_{1}\in%
\mathbb{C}
,c_{2}\geq0$ (we can always choose such decomposition). Now, taking a ket
state $\left\langle \psi\right\vert $ on $A\left\vert \psi\right\rangle ,$ we
get $\overline{A}=c_{1},$ and applying $\left\langle \psi\right\vert A$ on
$A\left\vert \psi\right\rangle $ yields $\left\langle \psi\right\vert
A^{2}\left\vert \psi\right\rangle =c_{1}^{\ast}c_{1}+c_{2}^{2}$, implying that
$c_{2}=\sqrt{\left\langle \psi\right\vert \left(  A-\overline{A}\right)
^{2}\left\vert \psi\right\rangle }.$

Without loss of generality, we consider the total operator acting on the
sample by%
\begin{equation}
A_{tot}=\sum_{i=1}^{N}A_{i}. \label{A_tot}%
\end{equation}
Suppose that we wish to measure the sample of particles followed by the total
operator $A_{tot}.$ For a measuring device followed by the von-Neumann
measurement, we have the following\ interaction Hamiltonian%
\begin{equation}
H=\lambda QA_{tot} \label{H_int}%
\end{equation}
for some coupling constant $\lambda>0,$ where $Q$ is\ conjugate to the pointer
$\Pi$ of the measuring device. We consider the evolution operator
$U=e^{-i\lambda QA_{tot}\Delta t}$ that acts on the coupled state $\left\vert
\Psi_{\text{total}}\right\rangle \left\vert \Pi\right\rangle $ where
$\left\vert \Pi\right\rangle $ is the pointer state of the measuring device.
Following the formula (\ref{formula}), the evolution of $\left\vert
\Psi_{\text{total}}\right\rangle $ is given by the superposition of two
orthogonal states%
\[
U\left\vert \Psi_{\text{total}}\right\rangle \left\vert \Pi\right\rangle
=\overline{U}\left\vert \Psi_{\text{total}}\right\rangle \left\vert
\Pi\right\rangle +\Delta U\left\vert \Psi_{\bot\text{total}}\right\rangle
\left\vert \Pi\right\rangle ,
\]
where $\left\vert \Psi_{\bot\text{total}}\right\rangle =\sum_{r=1}^{N}%
{\displaystyle\prod\limits_{i\neq r}^{N}}
\left\vert \psi\right\rangle _{i}\left\vert \psi_{\bot}\right\rangle _{r}$ is
orthogonal to $\left\vert \Psi_{\text{total}}\right\rangle .$ We note that the
above equation can be proved in a similar way to (\ref{formula}) by plugging
in $\left\langle \Psi_{\text{total}}\right\vert $ from both sides of the
equation to obtain the first part of the RHS, and then applying $\left\langle
\Psi_{\text{total}}\right\vert U$ to obtain the second part.

We define our time interval by $\Delta t=\tau/N$\ for some constant $\tau>0.$
By expanding the exponential operator $U$\ with respect to $\lambda,$ we
obtain%
\begin{align}
\overline{U}  &  =e^{-i\lambda Q\overline{A_{tot}}\tau/N}+\frac{1}{2}%
Q^{2}\Delta A_{tot}^{2}\left(  \frac{\tau}{N}\right)  ^{2}+...\label{Eq11}\\
&  =e^{-i\lambda Q\overline{A_{tot}}\tau/N}+O\left(  \frac{1}{N}\right)
,\nonumber
\end{align}
and%
\begin{equation}
\left\vert \Delta U\right\vert ^{2}=Q^{2}\Delta A_{tot}^{2}\left(  \frac{\tau
}{N}\right)  ^{2}+...=O\left(  \frac{1}{N}\right)  , \label{Eq112}%
\end{equation}
where $\Delta A_{tot}^{2}=N\cdot\Delta A^{2}$ and thus, $\Delta A_{tot}$ grows
as $\sqrt{N}.$ Then, for a sufficiently short time, $\Delta t,$ the orthogonal
state $\Delta U\left\vert \Psi_{\bot\text{total}}\right\rangle $ is arbitrary
small, and we get%
\begin{equation}
U\left\vert \Psi_{\text{total}}\right\rangle \left\vert \Pi\right\rangle
\approx e^{-i\lambda\overline{A_{tot}}\Delta tQ}\left\vert \Psi_{\text{total}%
}\right\rangle \left\vert \Pi\right\rangle , \label{exp111}%
\end{equation}
which gives a shift in the\ pointer of the measuring device $\left\vert
\Pi\right\rangle \rightarrow\left\vert \Pi-\lambda\Delta tN\overline
{A}\right\rangle ,$ so the observed value of $\lambda A_{tot}\ $after\ a short
time $\Delta t$ is $\lambda N\overline{A}.$

The outcome of our measurement, given our pointer $\Pi$, cannot depend on the
type of measurement we will do at a later time since, otherwise, we will break
causality. Such measurements can be, for instance, measurements on the whole
of the sample at once or measurements on each of the particles in the sample.
Thus, we know for sure that no matter what type of measurement we will
perform, we still will get a shift of the pointer by $\lambda N\overline{A}.$

Suppose now that instead of applying $A_{tot}$ on the sample state, we apply
$N$ measurements, each measurement with the underlying operator $A_{i}$ for
each of the particles in the sample. These measurements lead to the classical
probability of outcomes $p_{j}=N_{j}/N$ where $N_{j}$ is the number of
outcomes $\alpha_{j}$ that has been detected in the microscopic measurements,
and the\ average is clearly $\sum_{j}p_{j}\cdot\alpha_{j}.$

Following the fact that for sufficiently small $\Delta t,$ $e^{-i\lambda
\overline{A_{tot}}\Delta tQ}$ describes the eigenvalue of $U,$ for a post-
selected state $\left\vert post\right\rangle ,$ we find that the pointer will
have the same shift, regardless of the type of post-selected measurements one
will perform,
\begin{equation}
\left\langle post|U|\Psi_{\text{total}}\right\rangle \left\vert \Pi
\right\rangle =\left\langle post|\Psi_{\text{total}}\right\rangle \left\vert
\Pi-\lambda\Delta tN\overline{A}\right\rangle ,
\end{equation}
this shows that no matter what the post- selected state is, the pointer will
have the same shift. From the consistency between the macroscopic measurement
with the underlying sample state and the microscopic measurements, the result
must yield the same value. We then obtain the following equality%
\begin{equation}
N\sum_{j}\left(  b_{j}^{\ast}b_{j}\right)  \cdot\alpha_{j}=\sum_{j}N_{j}%
\cdot\alpha_{j}, \label{Equ11}%
\end{equation}
with recalling that $\left\langle \psi\right\vert A\left\vert \psi
\right\rangle \equiv\sum_{j}\left(  b_{j}^{\ast}b_{j}\right)  \alpha_{j}.$
Dividing both sides by $N$, we have $\sum_{j}\left(  b_{j}^{\ast}b_{j}\right)
\cdot\alpha_{j}=\sum_{j}p_{j}\cdot\alpha_{j},$ which holds, in general,\ if
and only if%
\begin{equation}
p_{j}=b_{j}^{\ast}b_{j}, \label{paa}%
\end{equation}
and we obtain the Born rule in its most explicit form.

While our proof was based purely on mathematical arguments and assumptions, it
can be derived following an explicit physical requirement. We conclude the
paper by proposing the following\ additional Postulate to the Postulates of
quantum mechanics:

\textbf{Postulate.} \textit{For a short\ enough time }$\Delta t\ $between two
measurements\textit{, a property, }$\mathcal{P}$\textit{, of a particle, will
keep its values fixed.}

The above Postulate is consistent and in accordance with every standard
experimental setup of quantum systems, and we can say that it is a physical
principle of quantum systems. In fact, for a property $\mathcal{P}^{\prime}$
of a quantum system that does not satisfy the Postulate, we can say that
$\mathcal{P}^{\prime}$ is not a property of the quantum system (for example,
see \cite{9}). The above Postulate allows us to prove the Born rule, as shown
earlier in the paper, while rejecting other candidates that do not fulfill the
proposed physical requirement. As a final remark, we note that following the
physical requirement given in the proposed Postulate, our proof of the Born
rule holds for both bounded and unbounded operators $A$ with finite
uncertainty $\Delta A.$

\bigskip

\textbf{Statements and Declarations.}

The authors have no competing interests to declare that are relevant to the
content of this article. Data sharing not applicable to this article as no
datasets were generated or analysed during the current study.


\begin{thebibliography}{9}
\bibitem {9}An example of a violation of the Postulate. Consider the rectangle
wavefunction $\psi\left(  x\right)  =rect\left(  x\right)  $, which is a
legitimate normalized wavefunction. Taking the free Hamiltonian $p^{2}/2m,$ we
see that the wavefunction is spreading almost immediately on the entire region
of $x\in%
\mathbb{R}
.$ This type of a wavefunction does not follow the additional physical requirement.
\end{thebibliography}
\end{document}